# QED Plasma at Extremely High Temperature and Density

Samina Masood [1]

[1] Department of Physical and Applied Sciences; University of Houston Clear-Lake

* Correspondance:   masood@uhcl.edu ; Tel.: (optional; include country code; if there are multiple corresponding authors, add author initials) +1281-283-3781 (F.L.).

**Abstract:** We use the renormalization scheme of QED (Quantum Electrodynamics) in real-time formalism to calculate the effective parameters of the theory, indicating the existence of relativistic QED plasma at extremely high temperatures and extremely high densities. High-density plasma is found inside the stellar cores and high temperature QED plasma could only exist, right after the neutrino decoupling temperature in the early universe, before the nucleosynthesis is complete. Radiation couples with the medium through the vacuum polarization in a hot and dense medium. Calculating the vacuum polarization tensor in a medium, the effect of radiation on matter is investigated in such a medium. We explicitly compute the parameters of QED plasma such as plasma frequency, Debye shielding length and the propagation frequency in terms of temperature and density of the superhot and superdense medium, respectively. This study helps to understand short term existence of QED plasma in the superhot and superdense systems.

## 1. INTRODUCTION

QED coupling constant alpha has a fixed value in vacuum and does not depend on the properties of the medium because the electromagnetic waves are purely transverse in nature and the photon is massless in vacuum. It does not interact with the medium and its longitudinal component has to be zero to satisfy the requirement of gauge invariance of QED lagrangian. In the presence of suitable statistical conditions, electromagnetic signals interact with the medium due to vacuum polarization. The quanta of electromagnetic signals in the medium are called Plasmon and acquire dynamically generated mass due to its interaction with fermions. This dynamically generated mass of Plasmon is calculated as a function of QED coupling parameter alpha that varies with temperature and density of the medium [1-11]. Above the neutrino decoupling temperature, electrons were interacting with photons as well as neutrinos in the medium and QED is not the only theory that describe the dynamics of this system. However, QED works perfectly well below the neutrino decoupling temperature [11].

When the electromagnetic waves propagated in the early universe at extremely high temperatures, QED coupling constant became a function of temperature [2, 7] and the coupling of the propagating particles is modified due to its interaction with the medium. On the other hand, inside the superdense stellar cores, Plasmon mass depend on the chemical potential μ, and generated due to the dynamical interaction of Plasmon with the medium [4-7]. Renormalization scheme is used to compute the vacuum polarization in such a medium in the real-time formalism [1-3] and the effect of the interaction of radiation with fermions of the medium. Renormalization is a process of removal of singularities in gauge theories. Perturbation theory is needed to compute the radiative corrections to higher order processes in a medium. Order by order cancellation of singularities is required by KLN (Kinoshita-Lee and Nauenberg) theorem [12-14] to assure the finiteness of a theory at all orders of perturbative expansion. Interaction of electromagnetic radiation with the fermions of the medium through vacuum polarizations modifies the QED coupling, which when is sufficiently increased, the phase transition can take place in the medium and the ideal gas approximation will not be applicable. Interaction with the medium induce temperature dependence to the physically measurable parameters [15-19] of the theory, which continuously grow with temperature. It has already been shown that the QED coupling of electromagnetically interacting medium is associated with the dynamically generated plasma screening mass [4--10] of photon and affects the propagation of photons in this medium. The phase of such a medium is determined by computing the plasma frequency and Debye shielding length as a function of temperature, which correspond to the relativistic plasma in the early universe. We will study the effect of temperature and density and the creation of relativistic plasmas in the extreme cases of temperatures and densities.

The paper is organized as follows: Section 2 briefly describes the calculational scheme of real-time formalism. This scheme is used to analytically calculate the vacuum polarization tensor of QED in Section 3 and the parameters of the relativistic plasma in Section 4. Results of Section 3 and 4 are applied to understand the propagation of particles in the early universe and the superdense stellar objects in Section 5.

## 2. CALCULATIONAL SCHEME

Thermal effects are included simply by replacing vacuum propagators by the modified propagators in a heat bath. Imaginary time formalism works in Euclidean space and time is taken as an imaginary coordinate and so is the energy. However, the imaginary time formalism does not provide the proper framework of perturbative calculations and order by order calculation of renormalization constants is not straightforward. Renormalization scheme of QED in real-time formalism allows an order-by-order cancellation of singularities, which cannot be calculated in Euclidean space. The corresponding real-time formalism in Minkowski space restores [2] the gauge invariance of QED with thermal propagator with the appropriate choice of the rest frame of the heat bath. The photon propagator of vacuum theory is modified [3] as:

$$\frac{1}{k^2} \rightarrow \frac{1}{k^2} - 2\pi i \delta(k^2) n_B(k), \qquad (1)$$

and the photon distribution function of photons is given as:

$$n_B(k) = \frac{1}{e^{\beta k}-1} \qquad (1a)$$

Where $\beta$ is the inverse temperature, i.e; $1/T$ and $k_0$ is the energy of the propagating photon. The corresponding fermion propagators is replaced as:

$$\frac{1}{p^2-m^2} \rightarrow \left[\frac{1}{p^2-m^2} + \Gamma_F(p,\mu)\right] \qquad (2)$$

with

$$\Gamma_F(p,\beta,\mu) = 2\pi i \delta(p^2-m^2)[\theta(p_0)n_F(p,\mu) + \theta(-p_0)n_F(p,-\mu)] \qquad (2a)$$

and

$$n_F(p, \pm\mu) = \frac{1}{e^{\beta(E,\pm\mu)}+1} = \Sigma_n(-1)^n e^{-n\beta(E\pm\mu)} \qquad (2b)$$

with $p_0$ the energy of electron. The chemical potential µ is positive for electron and negative for positron if the concentration of electron and positron is same.

It is shown [15-19] that renormalization of QED in a hot medium of the early universe yields the self-mass (or self-energy) of electron as

$$\frac{\delta m}{m}(T,\mu) \approx \frac{\alpha\pi T^2}{3m^2}\left[1 - \frac{6}{\pi^2}c(m\beta,\mu)\right] + \frac{2\alpha}{\pi}\frac{T}{m}a(m\beta,\mu) - \frac{3\alpha}{\pi}b(m\beta,\mu). \qquad (3a)$$

With the wavefunction renormalization constant as

$$Z_2^{-1}(T,\mu) = Z_2^{-1}(T=0,\mu) - \frac{2\alpha}{\pi}\int\frac{dk}{k}n_B(k) - \frac{5\alpha}{\pi}b(m\beta,\mu) + \frac{\alpha}{\pi}\frac{T^2}{E^2}\frac{1}{v}\ln\frac{1-v}{1+v}\times$$

$$\left[c(m\beta,\mu) - \frac{\pi^2}{6} - \frac{m}{T}a(m\beta,\mu)\right] \qquad (3b)$$

The renormalization constant of electron charge [8] is:

$$Z_3(T,\mu) \approx 1 + \frac{\alpha T^2}{6m^2}\left[\frac{ma(m\beta,\mu)}{T} - c(m\beta,\mu) + \frac{1}{4}(m^2 + \frac{\omega^2}{3}b(m\beta,\mu)\right] \qquad (3c)$$

Now, $Z_3$ in Eq. (3c) corresponds to the QED coupling constant α, which is related to the electric charge e through the relation α = e²/4π in the natural units. Here Masood's functions $a_i(m\beta,\mu)$ are evaluated for extreme conditions so at extremely high temperature (Eqs. 4) and at extremely large chemical potential (Eqn.5), and are given as:

$$a(m\beta) = \ln(1 + e^{-m\beta}), \qquad (4a)$$

$$b(m\beta) = \sum_{n=1}^{\infty}(-1)^n \text{Ei}(-nm\beta), \qquad (4b)$$

$$c(m\beta) = \sum_{n=1}^{\infty}(-1)^n \frac{e^{-nm\beta}}{n^2}, \qquad (4c)$$

For extreme temperatures of the early universe and these functions attain the following forms for extremely large densities.

$$a(m\beta, -\mu) = \mu - m \qquad (5a)$$

$$b(m\beta, -\mu) = \ln(\mu/m) \qquad (5b)$$

$$c(m\beta, -\mu) = \frac{\mu^2 - m^2}{2} \qquad (5c)$$

$$d(m\beta, -\mu) = \frac{\mu^3 - m^3}{3} \qquad (5d)$$

All the physically measureable parameters of the theory are calculated in terms of these ABC functions, first defined by Masood, et.al in literature [4-9].

1. **Vacuum Polarization Tensor**

The vacuum polarization tensor $\Pi_{\mu\nu}$ is a 4×4 matrix, which describes the propagation of light in 4-dimensional space and explains the propagation of light in 3-dimensional space with time. Transversality of light is associated with the zero mass of photon as a gauge requirement. Masslessness of photons is related to the absence of interaction of light with the medium and leads to the absence of its longitudinal component. However, in its general form, $\Pi_{\mu\nu}$ is expressed in a 4-dimensional space as [2, 10]

$$\Pi_{\mu\nu}(K,\mu) = P_{\mu\nu}\Pi_T(K,\mu) + Q_{\mu\nu}\pi\Pi_L(K,\mu) \qquad (6a)$$

$\Pi_L(K,\mu)$ and $\Pi_T(K,\mu)$ correspond to the longitudinal and transverse components of the vacuum polarization tensor respectively. In an interacting fluid, the photon acquires the dynamically generated mass, which adds up to a nonzero longitudinal component, and the photons with nonzero screening mass are not quanta of a transverse wave and has the ability to interact with the medium. Particle properties of photons help to understand the electromagnetic properties of such a medium. The vacuum polarization tensor is expressed in terms of longitudinal and transverse components in Eq. (5b).

$$\Pi_{\mu\nu} = \begin{pmatrix} -\frac{k^2}{K^2}\pi_L & -\frac{i\omega k_1}{K^2}\pi_L & -\frac{i\omega k_2}{K^2}\pi_L & -\frac{i\omega k_3}{K^2}\pi_L \\ -\frac{i\omega k_1}{K^2}\pi_L & \left(-1-\frac{k_1^2}{k^2}\right)\pi_T + \left(\frac{\omega^2 k_1^2}{k^2 K^2}\right)\pi_L & \left(-\frac{k_1 k_2}{k^2}\right)\pi_T + \left(\frac{\omega^2 k_1 k_2}{k^2 K^2}\right)\pi_L & \left(-\frac{k_1 k_3}{k^2}\right)\pi_T + \left(\frac{\omega^2 k_1 k_3}{k^2 K^2}\right)\pi_L \\ -\frac{i\omega k_2}{K^2}\pi_L & \left(-\frac{k_1 k_2}{k^2}\right)\pi_T + \left(\frac{\omega^2 k_1 k_2}{k^2 K^2}\right)\pi_L & \left(-1-\frac{k_2^2}{k^2}\right)\pi_T + \left(\frac{\omega^2 k_2^2}{k^2 K^2}\right)\pi_L & \left(-\frac{k_2 k_3}{k^2}\right)\pi_T + \left(\frac{\omega^2 k_2 k_3}{k^2 K^2}\right)\pi_L \\ -\frac{i\omega k_3}{K^2}\pi_L & \left(-\frac{k_1 k_3}{k^2}\right)\pi_T + \left(\frac{\omega^2 k_1 k_3}{k^2 K^2}\right)\pi_L & \left(-\frac{k_2 k_3}{k^2}\right)\pi_T + \left(\frac{\omega^2 k_2 k_3}{k^2 K^2}\right)\pi_L & \left(-1-\frac{k_3^2}{k^2}\right)\pi_T + \left(\frac{\omega^2 k_3^2}{k^2 K^2}\right)\pi_L \end{pmatrix}$$

................... (6b)

It shows that how the longitudinal component let the wave grow or die with time at the same point with the increase or decrease in the longitudinal component. $\Pi_{0i}$ or $\Pi_{i0}$ and even $\Pi_{00}$ components of the vacuum polarization vanish if the longitudinal component vanishes. Energy change associated with the change in the longitudinal component indicates that the frequency and wavelength of electromagnetic signal will change with time depending on the change in $\Pi_L$ and $\Pi_T$ and it will affect the velocity of propagation as well. Other components of the vacuum polarization tensor are also affected with nonzero $\Pi_L$ indicating the change in distribution of signal in space and the modification in the polarization properties with temperature. Vacuum polarization tensor $\Pi_{\mu\nu}$ plays a key role in computing the electromagnetic properties of the medium itself and leads to the calculation of plasma generating mass and then the Debye shielding length, indicating the phase change in to the relativistic plasmas. It can easily be seen if $\Pi_L=0$, the transverse wave will look like Eq. (6a)

$$P_{\mu\nu} = \begin{pmatrix} 0 & 0 & 0 & 0 \\ 0 & -1-\frac{k_1^2}{k^2} & -\frac{k_1 k_2}{k^2} & -\frac{k_1 k_3}{k^2} \\ 0 & -\frac{k_1 k_2}{k^2} & -1-\frac{k_2^2}{k^2} & -\frac{k_2 k_3}{k^2} \\ 0 & -\frac{k_1 k_3}{k^2} & -\frac{k_2 k_3}{k^2} & -1-\frac{k_3^2}{k^2} \end{pmatrix} \qquad (6c)$$

And if the electromagnetic signal were just the longitudinal with $\Pi_T=0$, it will look like

$$Q_{\mu\nu} = \begin{pmatrix} -\frac{k^2}{K^2} & -\frac{i\omega k_1}{K^2} & -\frac{i\omega k_2}{K^2} & -\frac{i\omega k_3}{K^2} \\ -\frac{i\omega k_1}{K^2} & \frac{\omega^2 k_1^2}{k^2 K^2} & \frac{\omega^2 k_1 k_2}{k^2 K^2} & \frac{\omega^2 k_1 k_3}{k^2 K^2} \\ -\frac{i\omega k_2}{K^2} & \frac{\omega^2 k_1 k_2}{k^2 K^2} & \frac{\omega^2 k_2^2}{k^2 K^2} & \frac{\omega^2 k_2 k_3}{k^2 K^2} \\ -\frac{i\omega k_3}{K^2} & \frac{\omega^2 k_1 k_3}{k^2 K^2} & \frac{\omega^2 k_2 k_3}{k^2 K^2} & \frac{\omega^2 k_3^2}{k^2 K^2} \end{pmatrix} \quad (6d)$$

In the non-interacting fluids $\Pi_L=0$ and $\Pi_{\mu\nu}(K,\mu) = P_{\mu\nu}\Pi_T(K,\mu)$. Therefore, nonzero value of $\Pi_L$ is a measure of deviation from the transverse behavior in an interacting fluid. If it is ignorable or sufficiently small, the fluid will affect the speed and have a small longitudinal component whereas it will have a sizeable effect for large values of $\Pi_L$. When the neutrino decoupling takes place, enough neutrinos are present in the medium to interact with the virtual electrons, which are produced during the vacuum polarization. Interaction of the neutrinos with these electrons give real contribution and the properties of such a medium are significantly modified as photon generates electrons which couple with neutrinos weakly and lead to electron-neutrino scattering processes. Interaction of virtual electrons with the magnetic field (or the photon in regular fluids or Plasmon in interacting fluids) also contribute to the magnetic moment of neutrino [20-24] which has already been shown to vary with temperature and density and may also related to the refractive energy of the medium [22].

The longitudinal and transverse components of the vacuum polarization tensor $\Pi_L(K,\mu)$ and $\Pi_T(K,\mu)$ respectively are evaluated in literature as:

$$\pi_L \cong \frac{4e^2}{\pi^2}\left(1-\frac{\omega^2}{k^2}\right)\left[\left(1-\frac{\omega}{2k}ln\frac{\omega+k}{\omega-k}\right)\left(\frac{ma(m\beta,\mu)}{\beta}-\frac{c(m\beta,\mu)}{\beta^2}\right)+\frac{1}{4}\left(2m^2-\omega^2+\frac{11k^2+37\omega^2}{72}\right)b(m\beta,\mu)\right]$$
(7a)

and

$$\pi_T \cong \frac{2e^2}{\pi^2}\left[\left\{\frac{\omega^2}{k^2}+\left(1-\frac{\omega^2}{k^2}\right)ln\frac{\omega+k}{\omega-k}\right\}\left(\frac{ma(m\beta,\mu)}{\beta}-\frac{c(m\beta,\mu)}{\beta^2}\right)+\frac{1}{8}\left(2m^2+\omega^2+\frac{107\omega^2-131k^2}{72}\right)b(m\beta,\mu)\right]$$
(7b)

This information of temperature dependence of the longitudinal and transverse components of the vacuum polarization tensor help to explicitly understand the electromagnetic properties of a medium at extremely high temperatures as well as the phase transition from an ideal gas to interacting fluid and relativistic plasma with the rise of temperature.

## 2. PARAMETRS OF QED PLASMA

The renormalization constants of QED in hot and dense medium can be described as the effective parameters of QED in a QED plasma. Photon acquire plasma screening mass and affect the coupling constant which changes the electromagnetic properties of the medium. It is shown [4-9] that the photons in this medium develop a plasma screening mass which can be obtained from the longitudinal and transverse component of the vacuum polarization tensor $\Pi_L(0,k)$ and $\Pi_T(k,k)$ where $K^2 = \omega^2 - k^2 = 0$ in vacuum [2, 9] and $\omega^2 = k^2$. This transversality condition changes with temperature and lead to the phase transition in a medium.

Longitudinal and transverse components ($\Pi_L$ and $\Pi_T$, respectively) of the vacuum polarization tensor $\Pi_{\mu\nu}$ play a crucial role in the determination of the electromagnetic properties of such a medium. Parameters such as propagation speed $v_{prop}$, refractive index $i_r$, dielectric constant, electric permittivity $\varepsilon(K)$, magnetic permeability $\mu(K)$, and the magnetic moment $\mu_a$ of electromagnetically interacting particles, propagating in the medium are studied in detail to determine the behavior of the medium. All of these parameters are expressed in terms of the temperature and density dependent values of $\Pi_L$ and $\Pi_T$ [2, 9], given in Eqs. (7) as:

$$\varepsilon(K) = 1 - \frac{\Pi_L}{K^2}, \qquad (8a)$$

$$\frac{1}{\mu(K)} = 1 + \frac{K^2 \Pi_T - \omega^2 \Pi_L}{k^2 K^2}, \qquad (8b)$$

and [6],

$$\varepsilon(K) = 1 + \chi_e \qquad (8c)$$

$$\mu(K) = 1 + \chi_m \qquad (8d)$$

Whereas, $\chi_e$ and $\chi_m$ give the dielectric constant and magnetization of a medium at given temperature and chemical potential, respectively. The relative change in the refractive index n with K is evaluated in terms of the parameters of the medium. The inverse of the square root of the product of magnetic permeability and electric permittivity correspond to the speed of propagation of the signal, which is set equal to 1 in vacuum due to the constant speed of light. The longitudinal and transverse components is evaluated from the vacuum polarization tensor directly, for all relevant ranges of the photon frequency $\omega$ and the wavenumber k. The propagation of particles in a medium is affected by the interaction of the particles with the particles around it and depend on the electromagnetic properties of the background medium. In free space, the speed of light is expressed in terms of the free space parameters of electric permittivity $\varepsilon_0$ and the magnetic permeability $\mu_0$ as:

$$c = \sqrt{\frac{1}{\varepsilon_0(K)\mu_0(K)}}$$

Substituting the values of the medium contribute to the refractive index of the medium as:

$$n(K) = \sqrt{\mu\varepsilon}$$

For a purely transverse signal, the refractive index is related to the opacity of a medium as the light can be blocked and delayed. However, the longitudinal frequency and the wavelength is zero due to the absence of interaction with the medium. At high temperatures of the early universe, excessive production of fermions lead to the sizeable thermal corrections to vacuum polarization, generating a nonzero and significantly large longitudinal component of the polarization tensor that could not be ignored. Thermal contribution to electrical permittivity (Eq. 8a) and the magnetic permeability $\mu(K)$. Eq. (8b) is used to evaluate thermal contribution to the propagation speed of electromagnetic waves and other relevant parameters in the early universe and it is given by,

$$v_{prop} = \sqrt{\frac{1}{\varepsilon(K)\mu(K)}} \qquad (9a)$$

which corresponds to the refractive index $r_i$ of the medium as:

$$r_i = \frac{c}{v} = \sqrt{\frac{\varepsilon(K)\mu(K)}{\varepsilon_0(K)\mu_0(K)}} \qquad (9b)$$

Longitudinal component of the vacuum polarization tensor at finite temperature is used to determine the phase of the medium and indicate overall properties of the medium for the relevant phase. Medium properties are changing with temperature due to modified electromagnetic couplings (3c). $K_L$ is evaluated by taking $\omega = k_0 = 0$ and p very small, the Debye shielding length $\lambda_D$ of such a medium is then given by the inverse of $K_L$

$$\lambda_D = 1/K_L \qquad (10a)$$

and, the corresponding frequency $\omega_D$ as

$$\omega_D = \frac{2\pi v_{prop}}{\lambda_D} = 2\pi K_L v_{prop} \qquad (10b)$$

Satisfying the relation

$$f_D \lambda_D = v_D \qquad (10c)$$

The plasma frequency $\omega_P$ is also related to Debye shielding frequency $v_D$ as

$$\omega_P = \frac{2\pi v_D}{\lambda_D} \qquad (11)$$

The longitudinal and transverse components of the photon frequency ω and the momentum k can be evaluated at different temperatures for given values of ω and k. However, the effective values are determined as

$$k = (k_L^2 + k_T^2)^{1/2} \qquad (12a)$$

Such that

$$\omega = (\omega_L^2 + \omega_T^2)^{1/2} \qquad (12b)$$

The plasma frequency is defined as $\omega_P^2 = \omega_T^2$ as $\omega_L^2 = 0$ and the Debye shielding length is obtained from the longitudinal component of wavelength using equation (10b). These results can easily be generalized to different situations using the initial values of $\Pi_T$ and $\Pi_L$ from Eq. (7) and substituting the relevant FTD values of ABC functions evaluated by Masood et.al.

3. RESULTS and DISCUSSIONS

Physically measureable values of QED parameters in the extremely hot universe, right after the big bang, becomes temperature dependent due to its interaction with the hot fermions of the universe [19]. Massless photons acquired dynamically generated mass in such a medium giving longitudinal component to the propagated waves. As a result, QED coupling became a slowly varying function of temperature [18]. This effect of temperature was associated with the energy and wavelength of radiation; and the discussion about the propagation of monochromatic light is more relevant. The presence of electrons in the extremely hot universe (at $T \geq 10^{10}$ K) is the major reason for providing thermal contribution to the electromagnetic properties of the system. Temperature measured in units of electron mass (0.511 MeV), that is approximately equal to the temperature $10^{10}$ Kelvin. Boltzmann constant $k_B$ is set equal to one in natural system of units. Electromagnetic parameters of vacuum, such as electric permittivity, magnetic permeability and dielectric constant are all normalized to unity in this system of units, for convenience. In real-time formalism, FTD contributions appear as additive contribution to vacuum values, especially at the one loop level. Thermal corrections appear as a temperature dependent additive term and can be analyzed independent of the corresponding vacuum values. Similarly density dependent contributions depend on the chemical potential of electrons and chemical potential is measured in units of electron mass, so we can set electron mass equal to one as well. It has been noticed that the significant additive density contributions are seen at high densities only.

Using the values of ABC functions of Masood, et.al; we can calculate the contribution of longitudinal and transverse components of vacuum polarization tensor and the corresponding components of electromagnetic signal frequencies and wavelengths from Eqns. (4) and (5), for the extreme conditions of temperature and density, respectively. Just to get an idea of thermal contribution, we plot the transverse and longitudinal components of the electromagnetic signals near nucleosynthesis temperatures using Eqns. (4) and the cores of superdense stars, using Eqns. (5).

The values of the longitudinal and transverse coordinates of the vacuum polarization tensor in different ranges of Plasmon energy are calculated in superhot systems, for two conditions, as:

(i) for $\omega \gg k$

$$\pi_L = -\frac{\omega^2 e^2 T^2}{3k^2} \qquad \pi_T = \frac{\omega^2 e^2 T^2}{6k^2} \qquad (13\ a)$$

And for (ii) $\omega \ll k$

$$\pi_L = \frac{e^2 T^2}{3} \qquad \pi_T = \frac{e^2 T^2}{6} \qquad (13\ b)$$

Using these value of Eq. (13), all the parameters of the theory can be calculated for the given ranges of energies as: needed for a particular system in certain conditions.

Fig.1 gives the plot of $k_L^2$ and $\omega_T^2$ as a function of temperature. It is clearly seen that for T, below the electron mass, thermal contribution is not significant and we can still use $\omega = k_0 = 0$, approximately. For larger T, these two functions, however, develop a quadratic dependence on temperature and start to decouple. There is a relatively slower growth in the transverse frequency as compared to the faster growth in longitudinal momentum k. The larger value of the longitudinal component of propagation vector $k_L$ indicate the increase in interaction of electromagnetic signals in the medium with the presence of electrons, as compared to the growth in transverse frequency $\omega_T$ with temperature. It shows the electromagnetic signals bend sharply in the small universe with slower increase in transverse oscillation and gives an explanation of signal trapping in the smaller and hotter universe. Fig. 1 gives a plot of thermal contributions to the QED signal properties around nucleosynthesis temperatures and below the neutrino decoupling temperature that is around 2MeV. It is also to be mentioned that T=3 m correspond to temperature around 1.5 MeV, that is close to nucleosynthesis temperature. This range is chosen to see how the effect of temperature becomes significant and the electromagnetically interacting medium may exist in plasma phase near nucleosynthesis. Computation of QED parameters and nonzero value of the plasma frequencies and Debye shielding lengths indicate the existence of QED plasma in the given range. Nonzero values of the longitudinal component indicate the existence of QED plasma. Since, pure QED plasma is not expected above the neutrino decoupling temperature. We limit our study around nucleosynthesis temperatures that is sufficiently below neutrino decoupling. Nucleosynthesis occur through beta decay or neutrino capture processes.

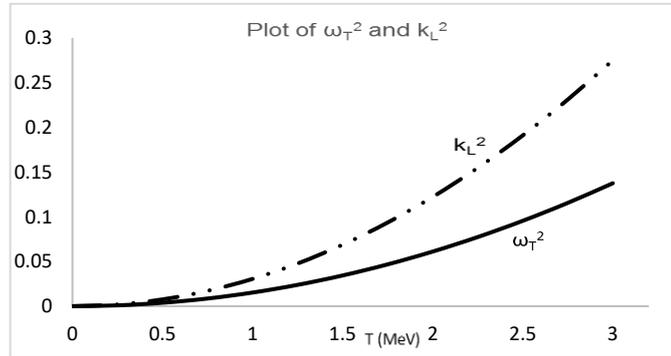

*Fig.1: Temperature dependence of the longitudinal component of the propagation vector (square) and the transverse frequency (square) are plotted showing that the Debye shielding length is greater than the plasma frequency.*

On the other hand, it is obvious from Fig. 2 that the transverse component of propagation vector is not much affected by temperature as it just depends on the b (mβ) and is ignorable as compared to quadratic function c(mβ). Moreover, the transversality of electromagnetic signals requires $k_L^2=0$ at temperature below the electron mass, because of the absence of the coupling with the medium. This coupling is not ignorable at T > $10^9$ K. Negligible change in $k_T^2$ gives the assurance that the transverse speed remains almost the same thus the $k_L^2$ has quadratic dependence on T, which indirectly affect the propagation velocity by mainly bending and then slowing it down inside the plasma, which ends up trapped inside plasma for particular frequencies. It seems to make perfect sense as the smaller volumes at higher densities and higher temperatures have more bending than the expanded volumes at lower temperatures, where the size of the universe was larger due to the expansion. These plots show how the plasma screening was dissolved with the creation of larger nuclei during nucleosynthesis, electrons and photons were not shielding each other when the nucleosynthesis was almost complete around temperature (T ≈ m).

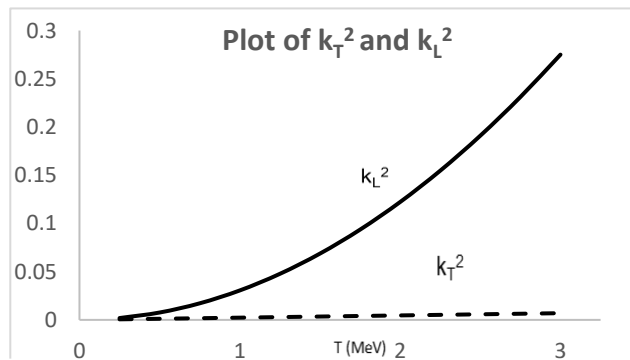

*Fig. 2: Comparing the plots of $k_L$ (square) and $k_T$ (square), it can be clearly seen that propagation vector in the transverse direction is independent of temperature. Temperature increases the inverse of Debye length ($1/k_L$) but transverse motion will not be affected.*

On the other hand high density effects are similar but they are more interesting for much higher densities which are 1-2 orders of magnitude larger than the electron mass. The superdense systems [25-29] are still under investigation and need much more observational data and testable working models for such systems. Since the information about superdense system is very limited, we cannot find the correct results without knowing the structure of superdense systems very well. However, it is not difficult to show from the above set of equations, that signals with particular frequency ranges can be trapped in superdense systems due to high refractive index or slow propagation. It is also clear from the ABC functions that the dominant contribution is a quadratic function of chemical potential.

At high densities, (i) we set $\omega = 0$, and $|k|$ approaches zero in this limit. In this limit we get

$$\pi_L(0,k) = K_L^2 = \frac{e^2 m^2}{2\pi^2}\left[\ln\frac{\mu}{m} + \frac{m^2}{\mu^2} - 1\right] \quad (14a)$$

$$\pi_T(0,k) = K_T^2 = \frac{e^2}{2\pi^2}\left[\frac{\mu^2}{8}\left(1 - \frac{m^2}{\mu^2}\right) - m^2 \ln\frac{\mu}{m} - \frac{1}{8}\left(\frac{m^2}{\mu^2} - 1\right)\frac{m^2}{4}\right] \quad (14b)$$

(ii) and in the limit, $\omega = |k|$ and both tend to zero, we get

$$\pi_L(|k|,k) = \omega_L^2 = 0 \quad \frac{e^2 m^2}{2\pi^2}\left[\ln\frac{\mu}{m} + \frac{m^2}{\mu^2} - 1\right] \quad (15a)$$

$$\pi_T(|k|,k) = \omega_T^2 = \frac{e^2}{2\pi^2}\left[\frac{\mu^2}{8}\left(1 - \frac{m^2}{\mu^2}\right) + m^2 \ln\frac{\mu}{m}\right] \quad (15b)$$

Eqs. (14 and 15) can be used to calculate all the parameters of an interacting fluid for the relevant densities of superdense media [27-31]. This study can be applied to understand the short term phases of matter in superhot and superdense systems of the early universe and compact starts, respectively. Moreover, if a frequency range is missing in the observational data, it may be an indication of internal trapping and not the absence of related situations. In that case, the missing frequencies may also be used to relate the statistical conditions and structure. Therefore, the calculation of the QED renormalization in superhot and superdense media is not only important to prove the physically acceptable behavior of QED in such media. It is worth investigating for the better understanding of the astrophysical objects as well. Theoretical understanding of the statistical background effect on different phases of matter can help to understand the astrophysical matter and the dynamics of such bodies at nanoscale level clearly.